\documentclass[%
 aip,
 amsmath,amssymb,
 reprint,%
]{revtex4-1}

\draft % marks overfull lines with a black rule on the right
\usepackage{CJKutf8}
\usepackage{graphicx}% Include figure files
\usepackage{dcolumn}% Align table columns on decimal point
\usepackage{xcolor}
\usepackage{bm}% bold math
\usepackage[utf8]{inputenc}
\usepackage[T1]{fontenc}
\usepackage{mathptmx}
\usepackage{etoolbox}

\makeatletter
\def\@email#1#2{%
 \endgroup
 \patchcmd{\titleblock@produce}
  {\frontmatter@RRAPformat}
  {\frontmatter@RRAPformat{\produce@RRAP{*#1\href{mailto:#2}{#2}}}\frontmatter@RRAPformat}
  {}{}
}%
\makeatother

\begin{document}
\begin{CJK*}{UTF8}{gbsn}
%\preprint{}

\title{Heterogeneous nucleation in the random field Ising model} %Title of paper



\author{Liheng Yao (姚立衡)}
%\email[]{ly343@cam.ac.uk}
%\thanks{}
\affiliation{DAMTP, Centre for Mathematical Sciences, University of Cambridge, Wilberforce Road, Cambridge CB3 0WA, United Kingdom}

\author{Robert L. Jack}
\email[]{rlj22@cam.ac.uk}
%\thanks{}
\affiliation{DAMTP, Centre for Mathematical Sciences, University of Cambridge, Wilberforce Road, Cambridge CB3 0WA, United Kingdom}
\affiliation{Yusuf Hamied Department of Chemistry, University of Cambridge, Lensfield Road, Cambridge CB2 1EW, United Kingdom}


\newcommand{\rlj}[1]{{\color{blue}#1}}
\newcommand{\ly}[1]{{\color{orange}#1}}

\date{\today}

\begin{abstract}
    We investigate the nucleation dynamics of the three-dimensional random field Ising model (RFIM) under an external field. We use umbrella sampling to compute the free-energy cost of a critical nucleus, and use forward flux sampling for the direct estimation of nucleation rates. For moderate to strong disorder, our results indicate that the size of the nucleating cluster is not a good reaction coordinate, contrary to the pure Ising model.  We rectify this problem by introducing a coordinate that also accounts for the location of the nucleus. Using the free energy barrier to predict the nucleation rate, we find reasonable agreement, although deviations become stronger as disorder increases. We attribute this effect to cluster shape fluctuations.  We also discuss finite-size effects on the nucleation rate.
    
    %We measure the free energy cost of cluster formation as a function of cluster size through umbrella sampling, and the distribution of committors at the peak of such free energy barriers. This analysis reveals that at low disorders the system behaves like the pure Ising model, whose nucleation dynamics is governed by classical nucleation theory (CNT). At higher disorders, although the transition still takes place through the nucleation and growth of a critical nucleus, the cluster size alone is no longer sufficient to describe the nucleation mechanics as the disorder causes nucleation to occur around a small number of preferred sites. We describe a procedure that predicts the location of such preferred sites from the disorder realization by sampling an ensemble of spherical nuclei. We then measure the free energy cost of nucleating clusters around such sites, and the committor distribution at the top of these barriers, and argue that the size of a cluster and the location of that cluster both need to be known to sufficiently describe the nucleation mechanics. We measure the nucleation rates using forward flux sampling (FFS), and show that they fit well with the lowest of the single-site nucleation barriers through an Arrhenius law. However, we observe that CNT increasingly overestimates the nucleation rate with increasing disorder due to shape fluctuations in the nucleating cluster. Finally, we analytically argue that due to disorder, the free energy barrier in our system decreases with system size $N$ by a term that asymptotically scales as $\sqrt{\log N}$.
\end{abstract}


\maketitle %\maketitle must follow title, authors, abstract and \pacs
\end{CJK*}

\section{Introduction}

Nucleation phenomena control many important physical processes, including vapor condensation \cite{mcdonald1962}, ice crystallization \cite{poole1992phase}, and many others \cite{oxtoby_homogeneous_1992, Sear2007, vsaric2014crucial, sosso2016crystal, arjun2019unbiased, bertolazzo2022polymorph, knopf2023atmospheric}.  Classical theories~\cite{volmer_keimbildung_1926, becker_kinetische_1935, zeldovich_theory_1943} describe the nucleation process as the spontaneous formation of a cluster of the stable product phase within a homogeneous metastable phase, across a free energy barrier that is the free energy cost of forming a critical nucleus. 
In modern formulations, this means that the reaction coordinate is the size of the nucleating cluster \cite{wolde_simulation_1996}. Classical nucleation theory (CNT) estimates this free energy barrier from macroscopic properties of the phases, and provides a good qualitative description of many nucleation processes \cite{oxtoby_homogeneous_1992}. In particular, nucleation in colloidal systems has been numerically studied in great detail \cite{wolde_simulation_1996, ten_wolde_computer_1998, Auer2001, auer_numerical_2004, auer_quantitative_2004}, establishing the modern computational approach to the study of nucleation dynamics. As a simple example of homogeneous nucleation, there is also an extensive body of numerical work on the domain-reversal dynamics of the Ising model~\cite{stauffer_monte_1982, acharyya_nucleation_1998, wonczak_confirmation_2000, brendel_nucleation_2005, ryu_validity_2010, cai_numerical_2010}, which is a simple and computationally tractable model system.

However, while these theories of homogeneous nucleation are elegant and consistent with computer simulation results, experimental systems are often affected by heterogeneous nucleation, for example due to random impurities, or surfaces.  These are beyond the scope of classical theories.
Studies of nucleation in Ising models have been extended to heterogeneous nucleation by manipulating the boundary conditions \cite{cirillo_metastability_1998, page_heterogeneous_2006, hedges_patterning_2012} or introducing impurities into the system \cite{sear_heterogeneous_2006, sear_non-self-averaging_2011, mandal_nucleation_2021}. In particular, it was found that nucleation in free boundary conditions happens preferentially at the corners of the system \cite{cirillo_metastability_1998}, and that the introduction of a single fixed spin can speed up nucleation by four orders of magnitude \cite{sear_heterogeneous_2006}. It was more recently found that the introduction of randomly placed $0$-spins lowers both the free energy barrier and the critical nucleus size \cite{mandal_nucleation_2021}.

In this work, we analyze nucleation in the random field Ising model (RFIM), which is a prototypical system for studying effects of disorder on first-order phase transitions.  The RFIM provides a schematic description of many physical systems where impurities play an important role, such as diluted antiferromagnets in a homogeneous external field \cite{fishman_random_1979}, mixed Jahn–Teller systems \cite{graham_random-field_1987}, binary liquids in porous media \cite{de_gennes_liquid-liquid_1984}, etc. (a review can be found in Ref.~\onlinecite{natterman_theory_1997}). As such, the model provides an interesting setting for effects of disorder on nucleation. Note for example that it interpolates smoothly between homogeneous nucleation when the disorder strength is zero, and heterogeneous nucleation at large disorder.
In addition, recent work has connected the RFIM to properties of glass-forming liquids~\cite{franz_glassy_2013, biroli2014, lin_scaling_2014, jack_phase_2016, ozawa_random_2018, biroli_random_2018, biroli_random-field_2018, guiselin_statistical_2022}, which provides a further motivation for studies of finite-temperature dynamics in the RFIM~\cite{roters_depinning_1999, roters_creep_2001, Dong_2012, Sinha2013, Mandal2014, yao2023}.

We study the three-dimensional RFIM using computer simulations.
An established approach for nucleation~\cite{cai_numerical_2010, mandal_nucleation_2021} is to use umbrella sampling \cite{torrie_monte_1974} to compute a free energy barrier associated with the critical nucleus, which can then be compared with direct estimation of the nucleation rate by forward flux sampling  (FFS)~\cite{allen_sampling_2005, allen_simulating_2006, allen_forward_2009} or other rare-event sampling methods~\cite{pan_dynamics_2004}.  Following the same path for the RFIM, an analysis based on the committor \cite{bolhuis_transition_2002} shows that that the size of the nucleating cluster is not a suitable reaction coordinate for nucleation, except when the disorder is very weak.  We rectify this problem by introducing a localized reaction coordinate, which measures the barrier for nucleation in a specific part of the system.  Using these results to predict nucleation rates, we compare with FFS simulations, finding agreement to within an order of magnitude over a wide range of nucleation rates, although deviations become stronger as disorder increases. We attribute these deviations to shape fluctuations of the nucleating cluster, which can affect the rate~\cite{pan_dynamics_2004,peters_reaction_2017}.  We also analyze finite-size effects on the nucleation rate, where rare regions of the system can play an important role.

In the remainder of this paper, Sec. \ref{the} introduces the RFIM and our theoretical approach. Sec. \ref{res} presents the results and Sec. \ref{con} shows our conclusions. We describe our numerical methods in the Appendices.

\section{Theory}
\label{the}

\subsection{RFIM}

We perform our investigation in three spacial dimensions, which is the lowest dimension at which the RFIM has a ferromagnetic phase \cite{imry_random-field_1975, binder_random-field_1983, aizenman_rounding_1989}. The three-dimensional RFIM is defined on a cubic lattice of linear size $L$. Each of the $N = L^{3}$ vertices on this lattice contains an Ising spin $s_i$ which takes the values $\pm1$, corresponding to the up-spin and down-spin states. A configuration of the system is denoted by $\bm{s}=(s_1,s_2,\dots,s_N$).

Each spin interacts with its nearest neighbors by an exchange interaction of strength $J > 0$, and feels a magnetic field of strength $H+h_i$, so the system's energy is
\begin{equation}
    E(\bm{s}) = -J \sum_{\langle ij \rangle}s_{i} s_{j} -\sum_{i}  (H+h_i) s_{i} \; ,
    \label{ham}
\end{equation}
where the notation $\langle ij \rangle$ indicates a sum over pairs of nearest neighbors, while the sum over $i$ run over all spins in the system. The parameter $H$ represents the external magnetic field, and $h_i$ is a quenched random field on site $i$ that are independent and identically distributed Gaussian random variables with standard deviation $R$. Physically, $R$ is the typical magnitude of the random field. We fix the energy scale by setting $J=1$, so the dimensionless parameters that appear in the energy are $H$ and $R$.

We study the domain reversal dynamics of the RFIM in a small positive external field, from the metastable state of bulk down spins to the stable state of bulk up spins. As the total number of up spins is not conserved, it is natural to use Metropolis dynamics \cite{metropolis_equation_1953}. In a single Monte Carlo (MC) move, one picks a random spin $s_i$ and proposes to change its value from $s_i$ to $-s_i$.  This proposed move is accepted with probability ${\rm min}(1,\exp \left(-\beta \Delta E_{i}\right) )$ where $\Delta E_{i}$ is the change in energy due to the proposed move. A sequence of $N$ such moves is called an MC sweep (MCS), which provides the natural time unit for our system.

\subsection{Becker-D{\"o}ring theory of nucleation}

Theories of nucleation aim to predict the nucleation rate per unit volume, denoted here by $I$.  In our case, this means that a system initialized in a metastable phase undergoes nucleation with probability $p_{\rm nuc}(\Delta t) = IN\Delta t$ in a short time $\Delta t$.  One typically expects that $I$ is an intensive quantity (independent of system size), which is the case for the pure Ising model.  However, situations may be more complicated in systems with disorder \cite{sear_non-self-averaging_2011}.

To estimate $I$, we start from Becker-D{\"o}ring theory \cite{becker_kinetische_1935, zeldovich_theory_1943}, which is framed in terms of the concentrations of clusters of up spins, and forms the basis of classical nucleation theory (CNT).  Write $M_n(\bm{s})$ for the number of up-spin clusters of size $n$ in configuration $\bm{s}$.  The free energy of such a cluster is measured relative to that of an individual monomer as
\begin{equation}
\beta F(n) = -\log \frac{\langle M_n\rangle}{\langle M_1\rangle}
\label{equ:Fn}
\end{equation}
We also identify $\rho_n=\langle M_n\rangle/N$ as the concentration of such clusters, and in particular $\rho_1$ is the concentration of isolated up spins.

Note that this theory does not distinguish the shapes of the clusters, nor their locations in the system.   Becker-D{\"o}ring theory assumes additionally that clusters grow and shrink by single spin flips, and that this process is Markovian.
The rates of growth and shrinkage are related through the detailed balance condition, expressed in terms of the equilibrium concentrations $\rho_n$~ \cite{frenkel_statistical_1939, ten_wolde_computer_1998}.

Finally one assumes that nucleation is a rare event in which case $F(n)$ will have a large barrier at a cluster size $n^*\gg 1$.  Then the size $n$ can be promoted to a continuous coordinate and the nucleation dynamics can be reduced to a one-dimensional Brownian motion in a potential $F(n)$. The nucleation rate is controlled by the barrier height as
\begin{equation}
    I_{\mathrm{BD}} = f^{*} \Gamma \rho_{1} e^{-\beta \Delta F} \; ,
    \label{arr}
\end{equation}
where $\Delta F = F(n^{*})$ and $f^{*}$ is the rate that the cluster size increases from $n^{*}$ to $n^{*} + 1$, and
\begin{equation}
    \Gamma = \left. \left( - \frac{1}{2 \pi} \frac{\partial^2 \beta F(n)}{\partial n^2} \right)^{\frac{1}{2}} \right|_{n = n^{*}}
    \label{zel}
\end{equation}
is the Zeldovich factor \cite{zeldovich_theory_1943} which gives {the extent to which the critical nucleus needs to grow before falling into the product basin}.

In CNT, the barrier height in \eqref{arr} is estimated in terms of macroscopic properties of the starting (metastable) phase and the nucleating (stable) one.  This is a drastic assumption since practical critical nuclei are unlikely to be macroscopic.  Fortunately, accurate microscopic computations of $\Delta F$ are possible using computer simulations~\cite{wolde_simulation_1996, ten_wolde_computer_1998, Auer2001, auer_numerical_2004, auer_quantitative_2004, pan_dynamics_2004, maibaum_phase_2008, cai_numerical_2010, mandal_nucleation_2021}.

\subsection{Reaction coordinate and committor}
\label{rea}

Modern theories for rare transitions between metastable states \cite{peters_reaction_2017} are framed in terms of a reaction coordinate, and describe the kinetic pathway by which the system transforms.  For nucleation, this pathway involves the growth of a cluster of the nucleating phase.  However, the most appropriate reaction coordinate for describing this process is a subtle question even for systems without disorder, involving an interplay of the cluster size and shape. For the RFIM, we show below that one must also consider the cluster location.

For consistency of Becker-D\"oring theory with modern rare-event theories, one should identify the reaction coordinate with the size of the nucleating cluster, which is the largest cluster of the stable phase in the system \cite{wolde_simulation_1996}.  In our system, this is the size of the largest connected cluster of up spins, which we denote by $\lambda(\bm{s})$. To test whether $\lambda$ is a good reaction coordinate, one should consider the committor $p_B$ \cite{bolhuis_transition_2002}: for any configuration $\bm{s}$, this $p_B(\bm{s})$ is defined as the probability that a trajectory initialized in $\bm{s}$ reaches the nucleating (stable) phase before it returns to the parent (metastable) one. This probability is estimated numerically by running many such trajectories.  %If $\lambda(\bm{s})$ is the true (optimal) reaction coordinate then $p_B(\bm{s})$ can be expressed as a function of $\lambda(\bm{s})$.

The ensemble of configurations $\bm{s}$ with $p_B(\bm{s})=0.5$ plays an important role in transition path theory: it is called the transition state ensemble (TSE).  If $\lambda$ is the optimal reaction coordinate, then the TSE can be characterized as the ensemble with $\lambda(\bm{s})=n^*$ where $n^*$ is the size of the critical nucleus.  This allows Becker-D\"oring theory to be interpreted in terms of this reaction coordinate. For the pure Ising model, this situation holds quite accurately for $\lambda$~\cite{cai_numerical_2010}. On the other hand, if a poor reaction coordinate is chosen, whose free energy maximum does not correspond to the TSE, one would expect faulty estimates of the reaction rate \cite{bolhuis_transition_2002, berezhkovskii_one-dimensional_2004}.

In addition to choosing an appropriate reaction coordinate, theories for homogeneous nucleation require some care because the nucleation rate also depends on the system size.
In standard rare event theories the rate is proportional to ${\rm e}^{-\beta\Delta F}$ and the free energy $F$ can be estimated via a histogram of the reaction coordinate, typically extracted by umbrella sampling.  However, it is important in nucleation theories that $\Delta F$ is instead computed via \eqref{equ:Fn}, see Ref.~\onlinecite{ten_wolde_computer_1998} and also Refs.~\onlinecite{maibaum_comment_2008, hedges_patterning_2012} for a discussion.  This free energy can still be computed from umbrella sampling simulations, see Appendix~\ref{appa} for details.

\section{Results}
\label{res}

\begin{figure}
    \centering
    \includegraphics[width=1.0\linewidth]{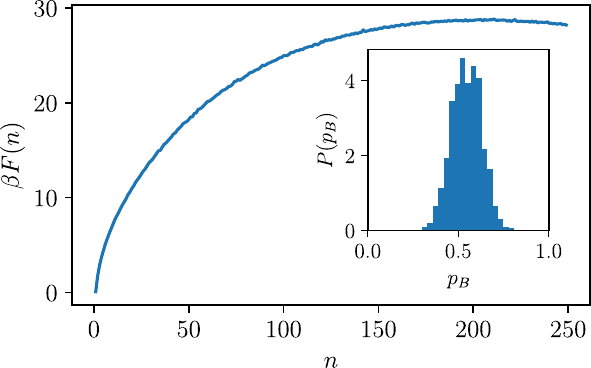}
    \caption{The free energy curve of a system with $R = 0.1$, $T = 2.7$ and $H = 0.45$. The inset shows the committor distribution of configurations taken at the top of the barrier. }
    \label{lowr}
\end{figure}

Analysis of nucleation requires a suitable choice of model parameters.  Writing $T_{c} (R)$ for the critical temperature of the RFIM, we must choose the temperature $T$ significantly below $T_c(R)$, but still high enough that simulations are tractable, as well as avoiding the roughening transition that occurs at low temperatures in the pure Ising model. We work throughout at $T = 2.7 \approx 0.6 T_{c} (0)$ which is a representative parameter choice within this regime. The external field $H$ must be chosen small enough that nucleation is a rare event, but very small values lead to very large critical nuclei, which are problematic for numerics. The values used in the following respect these constraints.

\subsection{Free energy and committor distribution}

For any given disorder realization the free energy $F(n)$ can be measured by umbrella sampling, which we describe in Appendix \ref{appa}.  An example is shown in Fig. \ref{lowr}, based on a representative realization with weak disorder $R=0.1$. Based on this free-energy profile, we estimate the size of the critical nucleus as $n^* \approx 210$ and we compute the committor distribution $P(p_B)$ for the ensemble with $\lambda(\bm{s})=n^*$. This is shown in the inset of Fig. \ref{lowr}.  The distribution shows a single peak near $p_B=0.5$, indicating that the cluster size is a suitable reaction coordinate for nucleation, for this (weak) disorder. (That is, configurations with $\lambda = n^*$ do form a good approximation for the TSE.) Comparing with $F(n)$ with results of Ref.~\onlinecite{cai_numerical_2010} for the pure Ising model, we observe that the addition of small disorder leaves the critical nucleus size almost unchanged, but slightly reduces the height of the barrier (the difference is approximately $5 k_{B} T$).  Reduced barrier heights are generic in the presence of disorder, as observed for example in heterogeneous nucleation \cite{sear_heterogeneous_2006, mandal_nucleation_2021}.

\begin{figure}
    \centering
    \includegraphics[width=1.0\linewidth]{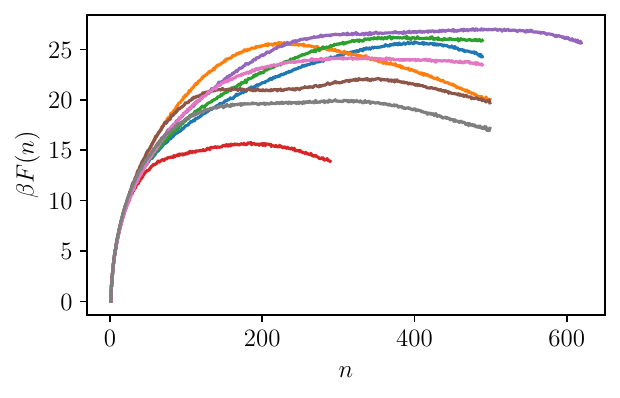}
    \caption{The free energy curves of eight different disorder realizations with $R = 1.0$, $T = 2.7$ and $H = 0.25$.}
    \label{midr}
\end{figure}

\begin{figure}
    \centering
    \includegraphics[width=1.0\linewidth]{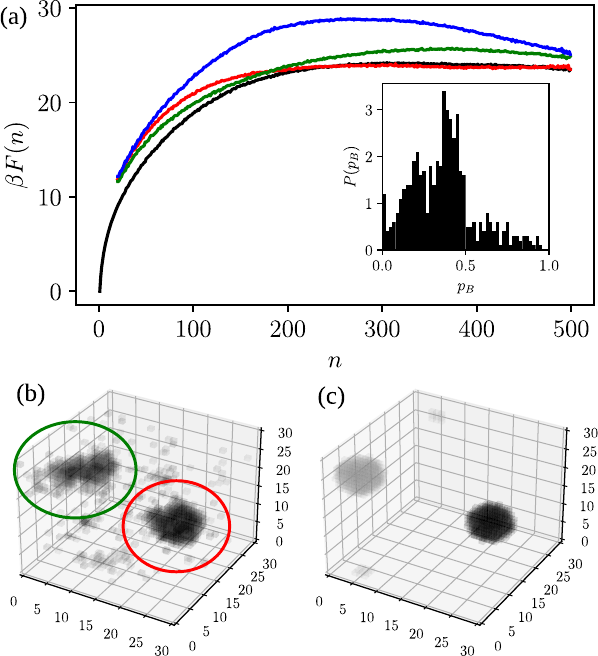}
    \caption{(a) The free energy curves obtained through umbrella sampling for a system with $R = 1.0$, $T = 2.7$ and $H = 0.25$. The black curve is the unconstrained free energy, while the colored curves represent free energies calculated using the constraint scheme introduced in Sec. \ref{local}. (b) The average local magnetization $m_{i}$ collected at the peak of the barrier, where darker patches corresponds to higher magnetization. The colors of the circles around the visible clusters match with the colors of their corresponding free energy curves. (c) The heatmap $W_{i}$ of the ensemble of spherical nuclei of size $n^*$, where darker patches corresponds to higher probability. The inset shows the committor distributions measured from configurations taken at the peak of the black free energy curve.}
    \label{umb}
\end{figure}

At higher disorder, the effect of the random field becomes much more pronounced.
%While at $R = 0.7$, the free energy curves are still convex in shape and cluster around the average (see Fig. \ref{size}), 
Fig. \ref{midr} shows free-energy profiles computed for 8 disorder realizations at $R = 1.0$. Their shapes vary significantly between disorder realizations, with some even showing nonconvexity, which can be attributed to  pinning: a set of sites with highly negative random field values around a growing cluster induces an energetic barrier to the growth of that cluster past these sites, which modifies the shape of the free energy curve. The behavior for intermediate disorder (R = 0.7) is qualitatively similar despite the free energy curves remaining convex, and will be discussed in Sec. \ref{sizes}, below.

Focusing on one of the realizations in Fig. \ref{midr}, $F(n)$ is replotted as the black curve in Fig. \ref{umb}(a). The corresponding committor distribution is shown in the inset of that Figure, showing that $P(p_B)$ is no longer sharply peaked, so the cluster size is not a good reaction coordinate, contrary to the assumption of classical theories, and that extra information is required to describe the nucleation mechanism.  This is due to the random field breaking translational invariance in the system: for a given cluster in the RFIM, the probability that it is expected to grow or shrink is not just a function of its size, but also heavily affected by the random field configuration around it.  We show in the following that a better reaction coordinate can be obtained by insisting that a cluster grows in a specific location.

\subsection{Localized reaction coordinate}
\label{local}

Our method for constructing an improved reaction coordinate is illustrated with the single disorder realization considered in Fig. \ref{umb}(a). We discuss the general case in at the end of this section.  For the the size $n^* \approx 340$ that maximizes $F(n)$, we extract representative configurations by umbrella sampling.  Fig. \ref{umb}(b) shows their average local magnetization $m_i=\langle s_i \rangle_{n^{*}}$, as a function of the position $i$.  [Here $\langle \cdot \rangle_{n^{*}}$ denotes the average over the ensemble with $\lambda(\bm{s}) = n^{*}$.]   One clearly sees a few locations where large clusters tend to appear.  This stands in stark contrast to the pure Ising model where translation invariance ensures that clusters are equally likely to form at any location, so $m_i$ would be independent of $i$.

For the RFIM we can therefore identify statistically preferred nucleation sites in the system by finding connected clusters of spins for which $m_i>m_{\rm cut}$, with a cutoff $m_{\rm cut}=-0.75$. Filtering out small clusters that represent background fluctuations, we index the resulting clusters by an index $j$, and write $C_{j}$ for the set of spins within the $j$th cluster. In the example of Fig.~\ref{umb}, we can easily identify two such preferred nucleation sites by visual inspection. For completeness, we also consider a third cluster which is not apparent from this visual representation, but does contribute strongly to the subset of configurations with large committor ($p_B>0.8$).

The resulting picture is that for a large system and a specific realization of the disorder, there are certain privileged locations where nucleation is most likely to occur.  This is attributable to local energy differences caused by the disorder. In other words, the transition still occurs by nucleation, but describing the nucleation mechanism requires analysis of the location of the nucleus, as well as its size. This is easily understood when one considers the pinning of cluster growth by the random field: for a cluster of a given size, the random field around its location dictates whether it is energetically favorable for the cluster to grow or shrink. For example, a small cluster at nucleation site A surrounded by a large number of positive random field spins may be more likely to grow and invade the system than a larger cluster at site B surrounded by highly negative random field spins.  Therefore the critical nucleus at site A will be smaller. Similarly, positive random field spins at the nucleation site make it energetically favorable for a cluster to form there, thus lowering the free energy barrier to cluster formation at that site, and vice versa.

Using this argument, we can predict the location of preferred nucleation sites from the random field realization alone by sampling the equilibrium ensemble of spherical nuclei of size $n^{*}$. We approximate a sphere of size $n_{\circ} \approx 340 = n^{*}$ centered around a spin $s_{i}$ by the set $O_{i}$ of all spins within a distance $r_{\circ} < 4.3$ from $s_{i}$. In the equilibrium ensemble of up-spin spheres of that size, a sphere centered at $s_{i}$ should appear with probability weight $w_{i} = \exp (\beta \sum_{j | s_{j} \in O_{i}} h_{j}) / Z_{\circ}$, where the sum runs over all spins within $O_{i}$, and $Z_{\circ}$ is a normalization factor such that $\sum_{i}^{N} w_{i} = 1$. A complete sampling of this ensemble can thus be performed by dropping one such sphere centered around each spin in the system, and calculating the probability that a spin $s_{i}$ is a member of a spherical nucleus of size $n^{*}$
\begin{equation}
    W_{i} = \sum_{j} w_{j} \theta (r_{\circ} - r(i, j)) \; ,
    \label{equ:Wi}
\end{equation}
where $r(i, j)$ is the distance between spins $s_{i}$ and $s_{j}$, and $\theta$ is the Heaviside step function. 

We plot the resultant configuration of $W_{i}$ in Fig. \ref{umb}(c). Despite the crudeness of this procedure, a comparison between Figs. \ref{umb}(b) and (c) shows that this ensemble of spherical nuclei predicts the actual preferred nucleation sites quite accurately. We however note that in this ensemble, the dark cluster in the bottom right corner of Fig. \ref{umb}(c) receives a probability weight of around 0.97, while the other cluster at the top left corner only receives a weight of around 0.02, which is different from the weights obtained from the actual sampling of the average local magnetization $m_{i}$. This is due to the crude assumption that all critical clusters are almost spherical in shape, see also Sec. \ref{rates}, below. Therefore, this procedure only estimates the locations of the nucleation sites, and is not sufficient to predict the probability that the system actually nucleates there. We also note that the preferred location of a spherical nucleus depends significantly on its size.

To make further progress, we estimate a nucleation rate $k_j$ associated with each preferred location $j$. This requires identification of a suitable reaction coordinate, and computation a suitable free energy barrier.  Then the nucleation rate $I$ for the whole system is obtained by summing over the rates for nucleation at each such location and dividing by the volume
\begin{equation}
    I=\frac{1}{N} \sum_j k_j \, .
\end{equation}
Our strategy in the following is to estimate individual contributions $k_j$ separately, and then to consider the total rate $I$.

The physical idea is that a suitable coordinate is the size $\lambda^{(j)}$ of the largest cluster in the vicinity of reference cluster $j$.
To achieve this, write $C^*(\bm{s})$ for the set of spins that forms the largest cluster in configuration ${\bm s}$ and let  $Q_{j}(\bm{s})$ be the number of spins in $C^*({\bm s})$ that overlap with the reference cluster $C_j$, that is $Q_{j}(\bm{s}) = |C^{*} (\bm{s}) \cap C_{j}|$.  Then define a localized reaction coordinate $\lambda^{(j)}(\bm{s}) = |C^{*}(\bm{s})|$ as the size of the largest cluster in ${\bm s}$, subject to the constraint that $Q_j$ is larger than a cutoff $D$, which we choose to be $\lambda^{(j)} / 6$. The choice of $D$ does not affect the measured free energy barrier as long as (a) it is smaller than the size of $C_{j}$, and (b) it ensures that there is significant overlap between $C^{*} (\bm{s})$ and $C_{j}$. Our particular choice is simply a matter of convenience.

We will see that this new reaction coordinate is suitable for identifying transition states {and measuring free energy barriers}, but we note that it does not make sense for small clusters (for example, no configuration can have $\lambda^{(j)}(\bm{s})<D$ according to this definition).

We then compute a free energy $F_j(n)$ along the reaction coordinate $\lambda^{(j)}$, which is directly comparable with the total free energy $F(n)$. (We emphasize again that we do not measure the free energy through a histogram of $\lambda^{(j)}$, recall Sec. \ref{rea}.)
For that purpose we define
\begin{equation}
    \chi_{j} (\bm{s}) = 
        \begin{cases}
        1, & \text{if } \lambda (\bm{s}) < n_{\mathrm{cut}} \text{ or } Q_{j} (\bm{s}) \geq D \\
        0,              & \text{otherwise}
        \end{cases}
    \; .
\end{equation}
Physically, $\chi_j=0$ if the largest cluster in the system is of size $n>n_{\rm cut}$ but \emph{not} at location $j$. The choice of $n_{\rm cut}$ does not affect the free energy barrier as long as $D$ follows the criteria mentioned above, and is sufficiently far away from any critical nucleus size. Here we choose $n_{\rm cut} = 20$.  Note that $\langle \chi_j \rangle$ is very close to unity because clusters bigger than $n_{\rm cut}$ are rare. 

Now define a constrained equilibrium distribution within which large clusters must be at location $j$:
\begin{equation}
p_{j}(\bm{s}) = \frac{1}{Z_{j}} \chi_j(\bm{s}) \exp\left[ -\beta E(\bm{s}) \right] \; ,
\end{equation}
where $Z_j$ is a normalization constant.  Averages with respect to this distribution are denoted by $\langle\cdot\rangle_j$.
Finally,  we define
\begin{equation}
\label{equ:Fj}
\beta F_j(n) = -\log \frac{ \langle M_n \rangle_{j}  }{ \langle M_1\rangle}  \; .
\end{equation}
By analogy with \eqref{equ:Fn}, this $F_j$ is an estimate of the free-energy profile associated with nucleation at  location $j$.
As advertised above, these profiles are directly comparable with the unconstrained profiles $F(n)$.  

The constraint $\chi_j$ means that $\langle M_n \rangle_{j}$ only counts large clusters when they are in location $j$.  For $n>n_{\rm cut}$ one may write
\begin{equation}
\label{equ:pj}
\langle M_n \rangle_j = p_j(n) \langle M_n \rangle
\end{equation}
where $p_j(n)$ is the probability that a cluster of size $n$ occurs at location $j$.  On the other hand, clusters with $n<n_{\rm cut}$ are almost unaffected by the constraint $\chi_j$ so one has $\langle M_n \rangle_j\approx \langle M_n \rangle$ in that case. 
This leads to a jump in $F_j$ at $n=n_{\rm cut}$ of size $\beta \Delta F_j^{\rm cut}=-\log p_j(n_{\rm cut})$.

The jump is clearly visible in numerical results for $\beta F_j$, shown as colored lines in Fig.~\ref{umb}(a), where the color of each curve indicates that the curve illustrates the free energy of cluster formation around the preferred nucleation site circled by the same color in Fig. \ref{umb}(b) (the blue curve corresponds to a site that is too faint to be seen in Fig. \ref{umb}(b), as discussed in the beginning of this Section). One also expects from (\ref{equ:Fj}) and (\ref{equ:pj}) that $F_j(n)\geq F(n)$: this bound is close to an equality if clusters of size $n$ in the unconstrained system occur predominately at location $j$.  This situation is realized for the red curve in Fig.~\ref{umb} in the range $200 \lesssim n \lesssim 400$.  

\begin{figure}
    \centering
    \includegraphics[width=1.0\linewidth]{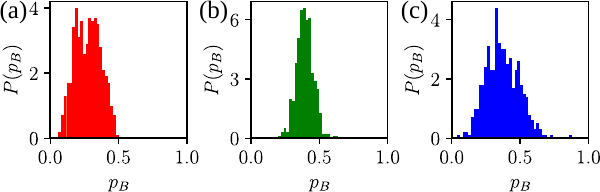}
    \caption{(a) - (c) The committor distributions measured from configurations taken at the peak of the colored free energy curves in Fig. \ref{umb}, with the colors of the histograms matching those of the free energies.}
    \label{com}
\end{figure}

To assess the suitability of $\lambda^{(j)}$ as a reaction coordinate, we identify the sizes $n_j^*$ of the critical nuclei at each location.  (Note that these sizes vary significantly between locations, from 260 to 370.) Then we extract configurations from the maxima of the three profiles $F_j$ from which we compute committor distributions $P(p_B)$.  These are shown in Fig.~\ref{com}.  Compared to the distribution in the inset of Fig.~\ref{umb}, they are much more sharply peaked, indicating that they successfully capture three different subsets of the transition state ensemble, associated with nucleation events at the three relevant locations.  Due to the flatness of the free energies near their peaks, it is difficult to locate the values of $n$ that yield committor distributions peaked very close to $p_{B} = 0.5$: the important observation is that the peaks are relatively narrow.

The rates for nucleation at each location can be estimated analogous to \eqref{arr} as 
\begin{equation}
    k_j = f^*_j \Gamma_j \langle M_{n^*_j} \rangle_j
    \label{equ:arr-k}
\end{equation}
where we used \eqref{equ:Fj} to express the relevant free energy barrier in terms of $\langle M_n \rangle_j$ (this is helpful because it shows that the jump in $F_j$ does not affect the rate estimates). See Sec.~\ref{rates} below for further discussion of these rates.

\begin{figure}
    \centering
    \includegraphics[width=1.0\linewidth]{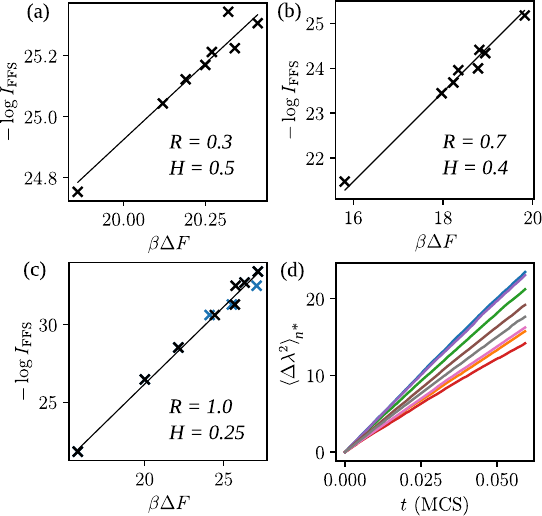}
    \caption{$-\log I_{\mathrm{FFS}}$ plotted against the free energy barrier, for $8$ disorder realizations at $T = 2.7$, (a) $R = 0.3$, $H = 0.5$, (b) $R = 0.7$, $H = 0.4$ and (b) $R = 1.0$, $H = 0.25$. The straight lines in both plots have gradient $1$. Then black crosses show free energy barriers calculated from unconstrained umbrella sampling runs, while the blue crosses in (c) show the true barriers calculated using the constraint scheme. (d) The ensemble-averaged time series $\langle \Delta \lambda^{2} \rangle_{n^*}$ plotted against $t$ at $R = 1.0$, $H = 0.25$, with each color corresponding to a different disorder realization.}
    \label{fit}
\end{figure}

The physical conclusion of this analysis -- and specifically of Fig.~\ref{com} -- is that domain reversal dynamics of the system is still controlled by the nucleation and growth of critical nuclei, which now takes place at preferred locations in the system. Our localized reaction coordinate accounts for this preference, and leads to single-peaked committor distributions. We repeated the above procedure for other disorder realizations and find similar results, though the shape and height of the barrier as well as the number of statistical dominant barriers vary between disorder realizations. The only exceptions to this behavior occur when neighboring target clusters are so close to each other that growth of one cluster occasionally invades the other, causing the committor distribution at the peak of the $F_{j}$ to lose its single-peaked shape.  This scenario is rare: we do not discuss it further here, but it should be straightforward to adapt the idea of a local reaction coordinate to this case, if required.

We also find that the free energy barriers extracted from $F_j$ are typically very similar to those obtained from $F$, as found in Fig.~\ref{umb}(a).  This may be expected from \eqref{equ:pj} as long as the number of locations for nucleation is not too large, such that $p_j=O(1)$.

\subsection{Nucleation rates and trajectories}
\label{rates}

In order to test the validity of (\ref{arr}) and (\ref{equ:arr-k}) in the RFIM, we calculate the nucleation rate in our system using forward flux sampling (FFS), which is described in Appendix \ref{appb}. The calculated rates are denoted as $I_{\mathrm{FFS}}$. We consider eight disorder realizations for three different values of $R$, and plot $-\log I_{\mathrm{FFS}}$ against the free energy barriers in Fig. \ref{fit}.  As free energy barriers calculated with or without the spacial constraint are shown to be extremely similar, we use the unconstrained scheme to calculate the values of $\Delta F$ unless stated otherwise. The results in in Fig. \ref{fit} fit well to a straight line of gradient $1$.  Moreover, the same straight line fit holds across disorder realizations for systems with the same parameters, implying that the kinetic prefactor in (\ref{arr}) (given by the intercept of the straight line) varies much more slowly than the exponential term for the same set of parameters, and can be treated as a constant across disorder realizations. This confirms that variations in nucleation rate is dominated by the variations of the free energy barrier.

We further comment that the fits in Fig. \ref{fit} cannot be used to distinguish between the our reaction coordinate and the conventional CNT one, as the barriers calculated with respect to the two coordinates are similar within the range of error acceptable to the fit. This is illustrated in Fig. \ref{fit}(c) where the data points using barriers calculated using the spatially constrained reaction coordinate are plotted in blue. The data points due to the two reaction coordinates either overlap completely or show only small variances. Instead, the quality of the reaction coordinates must be assessed by a committor analysis, as elaborated in the previous section.

\begin{figure}
    \centering
    \includegraphics[width=1.0\linewidth]{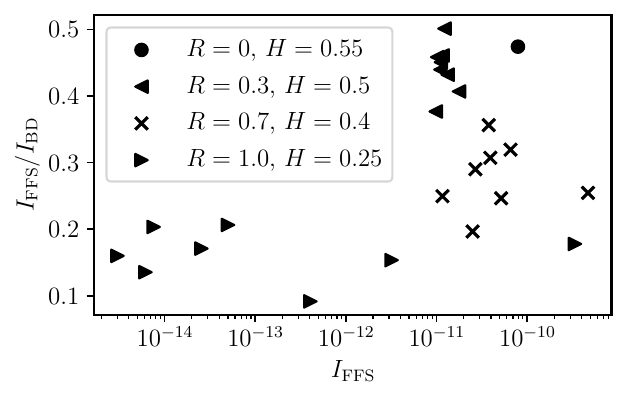}
    \caption{The ratio between the nucleation rates obtained from FFS and Becker-D{\"o}ring theory at $T = 2.7$ for different values of disorder and external field.}
    \label{ratio}
\end{figure}

We then compare predictions of (\ref{arr}) to the nucleation rates obtained by FFS. All terms on the right-hand side of (\ref{arr}) can be measured without explicitly measuring the nucleation rate: $\Gamma$, $\Delta F$ and $\rho_{1}$ can be obtained directly from the equilibrium cluster size distribution $\langle M_{n} \rangle$ through (\ref{zel}) and (\ref{equ:Fn}).  The parameter $f^{*}$ is computed in terms of the diffusion constant of the reaction coordinate at the top of the barrier. 
We can extract this diffusion constant by measuring the fluctuations of the reaction coordinate near the top of the barrier~\cite{peters_reaction_2017} through its mean-sqaure displacement
%mean square displacements of the reaction coordinate on top of the barrier, which is for short times linear in time as shown in Fig. \ref{fit}(d), so that $f^{*}$ is given by the constant gradient. 
\begin{equation}
    f^{*} = \lim_{t \to 0} \frac{\langle \Delta \lambda^{2} (t) \rangle_{n^{*}}}{2 t} \; ,
    \quad \Delta \lambda^{2} (t) = \left[ \lambda(t) - n^{*} \right]^2 \; ,
\end{equation}
where the average is taken over trajectories starting in the transition state ensemble (i.e., starting with $\lambda=n^*$). In practice, we collect 500 configurations at the peak of the barrier, and compute $\langle \Delta \lambda^{2} (t) \rangle_{n^{*}}$ by averaging over 200 trajectories starting from each configuration.  Results are shown in  Fig. \ref{fit}(d) for 8 disorder realizations at $R = 1.0$, and times $t$ up to $0.06$ MCS.  These results can be accurately fitted by straight lines, which we use to estimate $f^*$ through their gradients.

Combining all these results, we compare the rates predicted by \eqref{arr} with those measured using FFS, and plot their ratios $I_{\mathrm{FFS}} / I_{\mathrm{BD}}$ against the measured nucleation rate $I_{\mathrm{FFS}}$ in Fig. \ref{ratio} for a range of disorder strengths, varying the external field to ensure that the nucleation rates for all systems considered are comparable. We observe that while the nucleation rate varies by more than 4 orders of magnitude, the ratio of $I_{\rm BD}$ to $I_{\rm FFS}$ is always of order unity.  The theoretical prediction \eqref{arr} consistently overestimates the rate, by a factor between 2 and 10.

To interpret these results, we first note that result for the pure Ising model at $R = 0$ and $H = 0.55$ agrees with that reported by Cai and Ryu \cite{cai_numerical_2010}, who have produced a similar plot for the pure Ising model over a wide range of parameters.  Their results suggest that in three dimensions that \eqref{arr} systematically overestimates the nucleation rate by up to factor of 2.  For nonzero disorder, we make the following observations: (a) the nucleation rates increasingly vary between disorder realizations with increasing disorder strength. (b) For a given  set of parameters, the ratio $I_{\mathrm{FFS}} / I_{\mathrm{BD}}$ fluctuates weakly between disorder realizations, even when the nucleation rates vary by many orders of magnitude. This is particularly clear for $R = 1.0$. (c) The ratio $I_{\mathrm{FFS}} / I_{\mathrm{BD}} < 1$ for all parameter values, and decreases with increasing disorder strength $R$, showing that \eqref{arr} becomes less accurate when the disorder is strong. 

\begin{figure}
    \centering
    \includegraphics[width=1.0\linewidth]{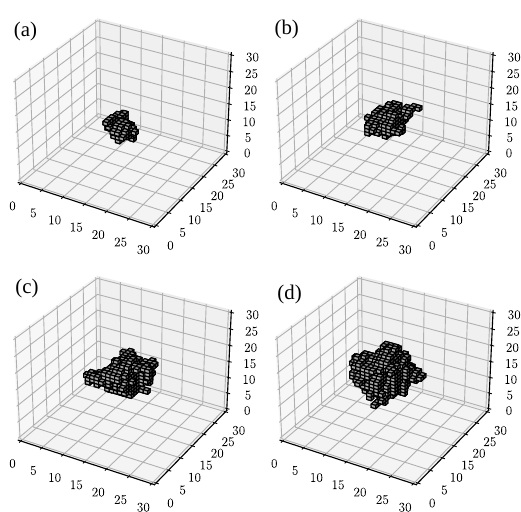}
    \caption{A sample nucleation trajectory extracted from a FFS run, showing the largest up-spin cluster in a system with $R = 1.0$, $T = 2.7$ and $H = 0.25$. The snapshots are taken at (a) $\lambda = 100$, (b) $\lambda = 200$, (c) $\lambda = 400$ and (d) $\lambda = 800$.}
    \label{traj}
\end{figure}

This last trend can be explained by the increasing importance of shape fluctuations of the nucleating cluster as disorder is increased. The nucleation rate predicted by \eqref{arr} is based on an effective coarse-grained one-dimensional description that integrates over all variables except for the size of the largest cluster.  (This includes an integration of cluster shape fluctuations.)  Such effective one-dimensional descriptions consistently overestimate the nucleation rate~\cite{berezhkovskii_one-dimensional_2004} unless the reaction coordinate is chosen to be exactly orthogonal to the surface on which all configurations have committor $0.5$. For the pure 3-dimensional Ising model, it was shown \cite{pan_dynamics_2004} that this surface is not orthogonal to the cluster size coordinate in a free energy landscape that is a function of the cluster size and cluster surface area, which implies that using cluster size as the only reaction coordinate will cause an overestimation of the nucleation rate.  This effect is particularly pronounced when shape fluctuations relax slowly, in comparison to microscopic time scales for addition or removal of single spins from the cluster.  As disorder increases, shape fluctuations will become slower and more significant, and the effects of neglecting them become more severe. Snapshots of the nucleating cluster taken from a sample FFS trajectory, as seen in Fig. \ref{traj}, show significant deviations from a spherical shape, which is linked to large slow fluctuations.

We also note that for the pure Ising model it was argued \cite{zia_effects_1985} that in the continuum limit shape fluctuations cause non-universal corrections to the nucleation rate in three dimensions but not two, which explains the better agreement between theory and experiment for nucleation rates in two dimensions that is reported in the literature \cite{cai_numerical_2010, mandal_nucleation_2021}.

\subsection{System size dependence of the nucleation rate}
\label{sizes}

\begin{figure}
    \centering
    \includegraphics[width=1.0\linewidth]{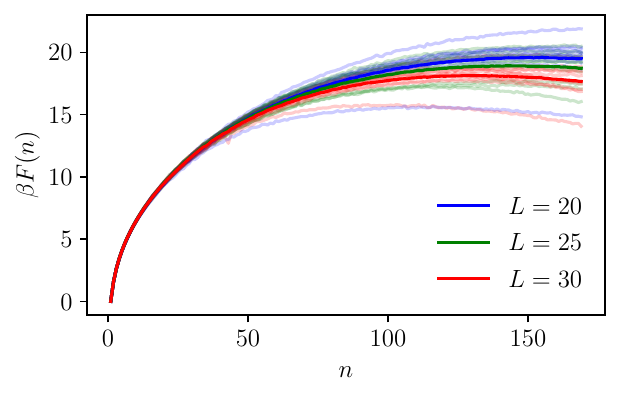}
    \caption{Free energy curves taken from systems of various sizes at $R = 0.7$, $T = 2.7$ and $H = 0.4$. The shaded lines represent free energies measured from $10$ disorder realizations, while the solid lines are averaged over those realizations.}
    \label{size}
\end{figure}

All numerical results thus far were obtained in systems of size $N=30^3$.
For systems without disorder then the probability of observing a (rare) nucleation event within a given (short) time window is an extensive quantity, hence our focus on the nucleation rate per unit volume, $I$.  The prediction \eqref{arr} is consistent with this observation because the concentrations $\rho_n$ are intensive quantities so the free energy barrier $\Delta F$ is independent of system size, in systems without disorder.

In systems with disorder -- like the RFIM -- the situation is more subtle~\cite{sear_statistical_2004}.  To illustrate this, Fig. \ref{size} shows the behavior of $F(n)$ as a function of system size, for several representative realizations of the disorder.  The sample-to-sample fluctuations are significant, but we observe a clear trend, that the average barrier height $\Delta F$ decreases as the system size $L$ increases.  The intuition for this result is that larger systems support a broader range of disordered local environments where nucleation can take place, and critical nuclei are biased towards (rare) regions where the random field happens to favor the nucleating phase.
This idea can be formalized using extreme value theory \cite{hansen2020, gumbel1958statistics}, which yields results similar to those of Sear \cite{sear_statistical_2004}, who considered a model of randomly distributed nucleation barriers.

Recalling \eqref{equ:Wi}, the main effect of the random field on critical nuclei is to reduce their energy by $\epsilon_O = \sum_{i\in O} h_i$, where $O$ is the set of spins that forms the nucleus. Taking $O$ to be a random cluster of size $n$, one sees that $\epsilon_O$ is Gaussian with mean zero and variance $nR^2$.  For a given disorder realization, it is useful to identify the location in the system where the nucleation barrier is smallest, since this will typically give the largest $k_j$ in \eqref{equ:arr-k}. This amounts to identifying the critical cluster $O$ with the largest value of $\epsilon_O$.

We denote this largest energy by $\epsilon_{\rm max}$.  Its behavior can be characterized within extreme value theory: suppose that all critical clusters have size $n$, and that there are  $a_n N$ possible cluster locations, and that each cluster has an independent value of $\epsilon_O$ (this last assumption is discussed in more detail below).  Then $\epsilon_{\rm max}$ is the largest of a large number of identically and independently distributed Gaussian random variables, so it has a Gumbel distribution whose most likely (modal) value is
\begin{equation}
    \epsilon_{\rm max}^* = R  \sqrt{n W \left(\frac{a_n^2  N^2}{2 \pi} \right)} \; ,
\end{equation} 
where $W$ is the Lambert $W$ function.  In fact, the probability density for $\epsilon_{\rm max}$ is
\begin{equation}
\begin{aligned}
    P(\epsilon_{\rm max}) &= \zeta \exp\left[ -\zeta(\epsilon_{\rm max}-\epsilon_{\rm max}^*) 
    - {\rm e}^{-\zeta(\epsilon_{\rm max}-\epsilon_{\rm max}^*)} \right] \; ,\\
     \zeta & = \frac{1}{R} \sqrt{\frac{1}{n} W\left(\frac{a_n^2 N^2}{2 \pi}\right)} \; .
    %\epsilon_{\rm max}^* &= R  \sqrt{n W \left(\frac{a_n^2}{2 \pi} N^2 \right)} \; ,
\end{aligned}
\end{equation}

The assumption that different clusters have independent values of $\epsilon_O$ is an approximation (due to possible overlaps) but for large enough systems one still expects to recover this limiting Gumbel distribution, with $a_n N$ playing the role of an effective sample size.

For large systems one uses that $W(x)\approx \log(x)$ at large $x$ to see that the typical value of $\epsilon_{\rm max}$ scales as 
\begin{equation}
    \epsilon_{\rm max}^* \approx R \sqrt{ 2n \log N} \; ,
    \label{equ:barrier-log}
\end{equation}
which increases (weakly) with $N$. That is, the lowest nucleation barrier in a large system decreases weakly with system size, contrary to the pure Ising model where it remains constant.  This is consistent with Fig.~\ref{size}

In the light of this result, one may imagine two scenarios.  Either nucleation is dominated in large systems by the cluster with largest $\epsilon_O$, so that a single term dominates the sum in \eqref{equ:arr-k}; or, there are many possible sites with similar barriers, which all contribute to the sum.  The weak $N$-dependence of \eqref{equ:barrier-log} means that the latter situation is realized in practice.  To see this, note that doubling the system size increases $\epsilon^*_{\rm max}$ by a small contribution of order $(\log N)^{-1/2}$ which carries through to the log-rate for nucleation; however, it also doubles the number of clusters with typical $\epsilon_O \approx R\sqrt{n}$, which corresponds to an increase of the log-rate by a finite constant $\log 2$. This latter contribution dominates the small contribution from the change in $\epsilon^*_{\rm max}$, leading to an extensive nucleation rate in large systems, albeit with strong finite-size corrections from clusters with anomalously large $\epsilon_O$.

\section{Conclusion}
\label{con}

We have investigated the nucleation dynamics of the three-dimensional RFIM by a combination of umbrella sampling and FFS. By calculating the distribution of committor probabilities at the peak of the free energy curves, we tested the hypothesis that the size of the nucleating cluster is a suitable reaction coordinate for nucleation. While this hypothesis is valid for weak disorder ($R \sim 0.1$), it breaks down at higher disorder ($R = 0.7$, $1.0$), where the location of the nucleating cluster is also needed to fully describe the nucleation dynamics.  We describe a method that predicts the locations of the preferred nucleation sites directly from the disorder configuration. Committor analysis confirms that a localized measure of cluster size serves as a good reaction coordinate, even in the presence of strong disorder.

We also find that while nucleation rates measured using FFS and our free energy barriers fit well through an Arrenhius form, Becker-D{\"o}ring theory increasingly overestimates the nucleation rates with increasing disorder. We attribute this behavior to the importance of shape fluctuations of the nucleating cluster, as such clusters observed in our simulations show highly anisotropic surfaces. Finally, we argue that the system size dependence of the nucleation rate manifests in a downward shift in the expected free energy barrier, proportional to $\sqrt{\log N}$.

Looking forward, a natural further step is to develop a theory that correctly accounts for the effects of shape fluctuations on nucleation rates. Previous attempts \cite{gunther_goldstone_1980, gunther_application_1994} have been made using Langer's theory of first passage times over a multidimensional landscape \cite{langer_statistical_1969} for systems without disorder. In addition, by assuming spherical critical nuclei, we have developed a procedure that predicts the location of the preferred nucleation sites directly from the disorder realization, but fails to predict their relative probability weights. A more accurate prediction of nucleation sites from the disorder will likely require machine learning methods.

Finally, we comment that the main results of this paper, namely the statistical preference of nucleation around a small number of locations determined by the disorder, the overestimation of the nucleation rate by CNT due to cluster shape fluctuations, and the nonlinear scaling of the nucleation rate with system size, should be generally applicable to nucleation in disordered environments.

\section*{Acknowledgments}

We thank Daan Frenkel for helpful discussions, and in particular for bringing Ref. \onlinecite{wolde_simulation_1996} to our attention.

\appendix
\section{Umbrella sampling}
\label{appa}

As nucleation is a rare event, sampling the ensemble average $\langle M_{n} \rangle$ is difficult for cluster sizes near the top of the barrier. We overcome this difficulty by using umbrella sampling \cite{torrie_monte_1974}, which we briefly describe here.

To force sampling in rare regions, a biasing potential is added to the energy, via the reaction coordinate. Since the underlying free energy is unknown, we simulate our system in multiple parallel windows, each with a harmonic biasing potential $\omega _{\alpha} (\lambda) = k (\lambda - \lambda_{\alpha})^2 / 2$, where $\lambda_{\alpha}$ are constants that determine the center of the biasing potentials, $k$ controls the strength of the biasing, and the index $\alpha$ runs through simulation windows. In practice, measuring $\lambda$ after each MC move is computationally expensive, so following Ref.~\onlinecite{wolde_simulation_1996}, we run a sequence of unbiased MC moves, and accept the entire sequence with a Metropolis rate ${\rm min}(1,\exp \left(-\beta \Delta \omega_{\alpha}\right) )$, where $\Delta \omega_{\alpha}$ is the change in the biasing potential due to the proposed sequence. To improve convergence, we also implement parallel tempering \cite{geyer_annealing_1995}, which exchanges configuration between simulation windows at a Metropolis rate ${\rm min}(1,\exp \left(-\beta \Delta \omega\right) )$, where $\Delta \omega$ is the change in the biasing potential due to the proposed exchange move. Following Ref.~\onlinecite{auer_quantitative_2004, auer_numerical_2004}, we exchange the center of the biasing potentials $\lambda_{\alpha}$ instead of the configurations between simulation windows. 

In practice, we collect 10000 samples of biased cluster size distributions $ M_{n} (\bm{s}_{k}^{\alpha})$ per window, where $\bm{s}_{k}^{\alpha}$ denotes the configuration of sample $k$ in window $\alpha$, and $0 < k < 10000$. We ensure convergence by checking and confirming that the histograms of $\lambda$ taken from neighboring windows overlap significantly, and the parallel tempering scheme has mixed the window indices sufficiently within the simulation time. The unbiased ensemble average is then estimated by the reweighting
\begin{equation}
    \langle M_{n} \rangle = \sum_{\alpha} \sum_{k} \Omega \left[ \lambda (\bm{s}_{k}^{\alpha}) \right] M_{n} (\bm{s}_{k}^{\alpha}) \; .
    %M_{n} = M_{n}^{\alpha} \exp \left[\beta \omega _{\alpha} (\lambda) \right] \left \langle \exp \left(-\beta \omega _{\alpha} (\lambda (\bm{s})) \right)\right\rangle
\end{equation}
The weights $\Omega$ are estimated using the unbinned weighted histogram analysis method (UWHAM) \cite{tan_theory_2012, varilly_uwham_2012}. $F(n)$ is then calculated from $\langle M_{n} \rangle$ through (\ref{equ:Fn}).

\section{Forward flux sampling}
\label{appb}

We measure the reaction rate and generate reaction trajectories using forward flux sampling (FFS) \cite{allen_sampling_2005, allen_simulating_2006, allen_forward_2009}. We define a set of interfaces $\lambda_{a}$ with $a=0,1,2,\dots,m$ and $\lambda_0<\cdots<\lambda_m$ in increasing order of the reaction coordinate $\lambda$. The interfaces are defined such that configurations in the reactant basin have $\lambda < \lambda_{0}$ and those in the product basin have $\lambda > \lambda_{m}$ ($\lambda_{m}$ is usually chosen to be much larger than $n^{*}$ to ensure that this is the case). The system is prepared with all spins pointing downwards, and allowed to evolve under Metropolis dynamics until it reaches the interface $\lambda_{0}$. This procedure is repeated until an ensemble of configurations with $\lambda_{0} < \lambda < \lambda_{1}$ is collected. 
The flux $I_{0}$ through the initial interface is then given by the number of collected configurations divided by the total time (in MCS) spent by the simulation in the reactant basin before reaching $\lambda_{0}$, summed over all collected configurations. 

A random configuration is then taken from this ensemble, and allowed to evolve under MC dynamics until it reaches $\lambda_{1}$ or returns to $\lambda_{0}$. Another ensemble of configuration with $\lambda_{1} < \lambda < \lambda_{2}$ is then collected, and the above steps repeated for each subsequent interface. The probability that a configuration collected at $\lambda_{a}$ reaches $\lambda_{a+1}$ before it returns to $\lambda_{0}$ can thus be calculated at each interface, and is denoted $P\left(\lambda_{a+1} \mid \lambda_a\right)$. As configurations with $\lambda > \lambda_{m}$ are considered to be in the product basin with probability $1$, the total nucleation rate is given by
\begin{equation}
    I_{\mathrm{FFS}} = \frac{I_{0}}{N} \prod_{a=0}^{m-1} P\left(\lambda_{a+1} \mid \lambda_a\right) \; .
\end{equation}

Using this method, one can generate trajectories from the reactant basin to the product basin. Moreover, the quantity
\begin{equation}
    p_{B}^{\mathrm{FFS}} (\lambda_{k}) = \prod_{a=m - 1}^{k} P\left(\lambda_{a+1} \mid \lambda_a\right) \;
\end{equation}
gives an estimate of the average committor of configurations at interface $k$.

% If in two-column mode, this environment will change to single-column format so that long equations can be displayed. 
%\begin{widetext}
%$$\mbox{put long equation here}$$
%\end{widetext}

% Figures should be put into the text as floats. 
% Use the graphics or graphicx packages (distributed with LaTeX2e).
% Put the label that you will use with \ref{} command in the braces of the \label{} command.
%
% \begin{figure}
% \end{figure}

% \begin{table}
% \end{tabular}
% \end{table}

%\begin{acknowledgments}
% Put your acknowledgments here.
%\end{acknowledgments}

\bibliography{bibliography}

\end{document}